\documentclass[11pt]{iopart}
\usepackage{graphics}                                                           
\usepackage{color}                                                              
\usepackage{ifthen}                                                             
\usepackage{mathrsfs}
\usepackage{amssymb}
\usepackage{bbold}
\usepackage{multirow}                                                          
\usepackage{tabularx}
\usepackage{bbm}
\newcommand\id{\ensuremath{\mathbbm{1}}} 
\def\nn{\nonumber}
\def\ds{{\rm d}s}
\def\da{{\rm d}a}
\def\dt{{\rm d}t}
\def\dx{{\rm d}x}
 
\begin{document}
%%%%%%%%%%%%%%%%%%%%%%%%%%%%%%%%%%%%%%%%%%%%%%%%%%%%%%%%%%%%%%%%%%%%%%%%%%%
\title{Tensor Minkowski Functionals for random fields on the sphere}

\author{Pravabati Chingangbam$^{1}$, K. P. Yogendran$^{2,3}$} 
\ead{prava@iiap.res.in, yogendran@iisertirupati.ac.in}
\author{Joby P. K.$^{1,4}$, Vidhya Ganesan$^{1,5}$}
\ead{joby@iiap.res.in, vidhya@iiap.res.in}
\author{Stephen Appleby$^{6}$, Changbom Park$^{6}$}
\ead{stephen@kias.re.kr, cbp@kias.re.kr}        
\address{$^1$ Indian Institute of Astrophysics, Koramangala II Block,       
  Bangalore  560 034, India\\                                                          
  $^2$ Indian Institute of Science Education and Research, Sector 81, Mohali,
  India\\                
$^3$ Indian Institute of Science Education and Research, C/o Sree Rama
Engineering College (Transit Campus), Karakambadi Road,                                 
Mangalam (P.O.), Tirupati 517 507, India\\
$^4$ Department of Physics, University of Calicut, Malappuram, Kerala-673 635, India\\
$^5$ Department of Physics, Indian Institute of Science, C. V. Raman Ave., Bangalore  560 012, India\\   
$^6$Korea Institute for Advanced Study, 85 Hoegiro, Dongdaemun-gu,              
Seoul 02455, Korea}  

%\date{\today}                                                                  
%%%%%%%%%%%%%%%%%%%%%%%%%%%%%%%%%%%%%%%%%%%%%%%%%%%%%%%%%%%%%%%%%%%             
\begin{abstract}                                                                
  We generalize the translation invariant tensor-valued Minkowski Functionals which are defined on two-dimensional flat space to 
  the unit sphere. We apply them to level sets of random fields. The contours enclosing boundaries of level sets of random fields give a spatial distribution of random smooth closed curves. We outline a method 
  to compute the tensor-valued Minkowski Functionals numerically for any random field on the sphere. 
  Then we obtain analytic expressions for the ensemble expectation values of the matrix elements for isotropic Gaussian and Rayleigh fields. The results hold on flat as well as any curved space with affine connection. We elucidate the way in which the matrix elements encode information about the Gaussian nature and statistical isotropy (or departure from isotropy) of the field. Finally, we apply the method to maps of the Galactic foreground emissions from the 2015 PLANCK data 
  and demonstrate their high level of statistical anisotropy and departure from Gaussianity. 
\end{abstract}
%\pacs{98.80.-k,98.80.Bp}

\maketitle
%%%%%%%%%%%%%%%%%%%%%%%%%%%%%%%%%%%%%%%%%%%%%%%%%%%%%%%%%%%%%%%%%%%%%%%%%%%
\section{Introduction} 

% Cosmological fields
Cosmological fields may be considered as random fields on two or three dimensional spaces. Our understanding of the broad features of the Universe are derived from statistical analysis of these fields. In two dimensions the most important cosmological fields are the Cosmic Microwave Background (CMB) temperature and polarization fields~\cite{Penzias:1965,Smoot:1992,Kovac:2002fg}.  
The theory of Gaussian random fields on the sphere relevant for the CMB fields was developed by Bond and Efstathiou~\cite{Bond:1987ub}, building on the excursion set theory of random fields developed by Adler~\cite{Adler:1981}.
    
% Scalar MFs
Scalar Minkowski Functionals (henceforth MFs) have been widely used in cosmology~\cite{Tomita:1986,Gott:1990,Mecke:1994,Schmalzing:1997,Schmalzing:1998,Winitzki:1998,Matsubara:2003yt}. (See also~\cite{Buchert:2017uup} for a comprehensive description of the development of the subject and list of references.) 
These are defined on real space and contain correlations of arbitrary order. They have been used in searches for non-Gaussianity of primordial origin~\cite{COBE_NG:2000,WMAP_NG:2011,Ganesan:2014lqa,Ade:2015ava,Planckiso:2015,Buchert:2017uup}, effect of residual foreground in CMB data~\cite{Chingangbam:2013} and lensing on CMB fields~\cite{Munshi:2016,Santos:2015}. They are however insensitive to morphological information related to the shape and relative alignment of structures. 
Quantities that are closely related to the scalar MFs and have also been applied to cosmological fields are the counts of hot and cold spots~\cite{Coles:1987,Chingangbam:2012,Park:2013} and extrema counts~\cite{Pogosyan:2009}. 

Scalar MFs are a subset of the wider class of tensor MFs~\cite{McMullen:1997,Alesker:1999, Beisbart:2002,Hug:2008,Schroder2D:2009,Schroder3D:2013} which have been defined for flat two- and three-dimensional spaces. Vector-valued MFs have been used to characterize galaxy morphologies~\cite{Beisbart:2001a,Beisbart:2001b}. In this paper we focus only on tensor-valued MFs. They can be classified into subsets of translation invariant and covariant tensors. 
In ~\cite{Vidhya:2016} Ganesan and Chingangbam introduced tensor-valued MFs for random fields defined on flat two-dimensional space and applied them to the CMB temperature and polarization fields. One of the translation invariant rank-2 MFs can capture the information about the shapes of individual hotspots and coldspots in the CMB fields (or any random field for that matter) and alignment in the spatial distribution of these structures. The method was applied to the PLANCK data released in 2015~\cite{Planckdata:2015} and it was shown that the $E$ mode data exhibit significant level of alignment. It is expected that the full PLANCK data, after it becomes publicly available, will shed more light on the physical origin of this alignment.

The approach in this work was numerical and has two important issues. The first is the use of stereographic projection of the CMB fields that are defined on the sphere to the plane. Since this projection is a conformal projection, angles remain invariant but sizes are scaled, which means that sizes of the closed iso-field contours get scaled. This can introduce numerical artefacts in the computation of the alignment of the contours. The second issue is that numerical errors associated with the pixellization increases with the level of anisotropy of the contours. These issues can be ameliorated if we compute the tensor-valued MFs directly on the sphere. For these reasons it is desirable to first generalize the definition of tensor-valued MFs to general smooth manifolds and then to have analytic handle on their understanding. This paper is a step towards that direction.

In this paper, we first review the definition of tensor-valued MFs in flat 2-dimensional space and focus on the translation invariant rank-2 MFs (henceforth TMFs). We discuss how one of them encodes information of the intrinsic anisotropy of structures (closed curves in two dimensions) and the relative alignment between many structures. We clarify that the intrinsic isotropy of a single structure as encoded in this particular TMF is just a manifestation of the $m$-fold rotational symmetry, with $m\ge 3$, of a closed curve. Further, for many structures we present a new geometric way of understanding statistical isotropy of the spatial distribution of structures. Next, we  
generalize the definition of the translation invariant TMFs to curved spaces, with emphasis on the unit sphere. Then we compute ensemble expectation values of the TMFs for isotropic Gaussian and Rayleigh random fields. We elucidate how the Gaussian/Rayleigh nature and the isotropy of the fields are encoded in the expressions of the TMFs. Then we describe the numerical implementation of the formulae for the TMFs on Gaussian isotropic fields. We further apply the method to maps of synchrotron, thermal dust, anomalous microwave and CO line emissions of ur Galaxy taken from the 2015 PLANCK data to demonstrate how the non-Gaussianity of these fields are captured by the elements of TMFs and the statistical anisotropy is captured by the alignment parameter. 
Earlier attempts to use the shapes of hotspots and coldspots of the CMB to get cosmological information can be found in~\cite{Gurzadyan:2003, Gurzadyan:2005, Aurich:2011, Berntsen:2012, Ajit:2011}.

This paper is organized as follows. In section 2 we review the definition of tensor Minkowski Functionals on flat 2-dimensional space and discuss the intrinsic anisotropy and relative alignment measures of structures. Then we generalize the definition of the rank-2 translation invariant TMFs to general smooth manifolds, specifically to the unit sphere and then discuss the measure of intrinsic anisotropy and alignment of random smooth curves on the sphere. In section 3 we apply the TMFs to random curves associated with level sets of random fields. Then we calculate analytically the ensemble expectation values of the TMFs for the special cases of isotropic Gaussian and Rayleigh fields. In section 4 we discuss the numerical computation of TMFs and apply to isotropic Gaussian CMB temperature maps, and maps of foreground emissions from the PLANCK data, specifically synchrotron, thermal dust, CO and anomalous microwave  emissions from our Galaxy.  We conclude with a discussion of the results, their implications and practical usefulness, and future applications in section 5.

%%%%%%%%%%%%%%%%%%%%%%%%%%%%%%%%%%%%%%%%%%%%%%%%%%%%%%%%%%%%%%%%%%%%%%%%%%%
\section{Tensor Minkowski Functionals on the unit sphere}

We begin by recapitulating the definition of general tensor MFs of
rank $(m,n)$ on flat 2-dimensional space. Then we focus on the
translation invariant rank-2 TMFs and generalize their definition to
curved space.

\subsection{Review of tensor Minkowski Functionals on flat 2-dimensional space}

Given a closed curve $C$, let $\vec r$ denote the position vector of a point on the curve, $\hat n$ denote the unit vector which is normal to the tangent vector, $\kappa$ denote the local curvature of the curve at the point. Using these quantities tensor MFs are defined as follows (see e.g. \cite{Schroder2D:2009}):
\begin{eqnarray} 
W_0^{m,0} &=&  \int_{C} \vec{r}^{\,m} \,\da \nonumber\\
W_1^{m,n} &=& \frac12 \int_{C} \vec{r}^{\,m} \otimes \hat{n}^{\,n}  \,\ds,\nonumber\\
W_2^{m,n} &=& \frac12 \int_{C} \vec{r}^{\,m} \otimes \hat{n}^{\,n} \,\kappa \,\ds,\
\label{eqn:tmf}
\end{eqnarray}
where $C$ denotes the closed curve, $\da$ is the area element of the region enclosed by the closed curve and $\ds$ is the infinitesimal arc length of the curve. The tensor product of two vectors is defined to be the symmetric product $(\vec A \otimes \vec B)_{ij}=\frac12 \left( A_iB_j+A_jB_i\right)$. $\vec r\,{}^m$ means  $m$-fold tensor product of $\vec r$, and similarly for $\hat n^n$. 

The rank-0 MFs are the usual scalar MFs. They differ from the usual expressions used in cosmology, (see for example~\cite{Schmalzing:1998}) by numerical factors. The rank-1 MFs are translation covariant. The set of tensor MFs of rank 2 can be subdivided into translation covariant and translation invariant ones. (See Table 1 of~\cite{Schroder2D:2009}). The translation invariant ones are $W_1^{1,1}$, $W_1^{0,2}$, $W_2^{1,1}$ and $W_2^{0,2}$.  Of these, $W_1^{0,2}$ and $W_2^{1,1}$ are dependent on each other. The three linearly independent translation invariant rank-2 TMFs maybe chosen to be
\begin{eqnarray}
    W_1^{1,1} &=& \frac12 \int_{C} \vec{r} \otimes \hat{n} \,\ds,\\
    W_2^{1,1} &=& \frac12 \int_{C} \vec{r} \otimes \hat{n} \,\kappa \,\ds,\\
      W_2^{0,2} &=& \frac12 \int_{C} \hat{n} \otimes \hat{n} \,\kappa \,\ds.
    \label{eqn:rank2}
\end{eqnarray}
They are related to the rank-2 tensors $W_j{\mathbf E}$, where $j=0,1,2$ and ${\mathbf E}\equiv \hat {\mathbf e}_1\otimes \hat{\mathbf e}_1+ \hat{\mathbf e}_2\otimes \hat{\mathbf e}_2$ is the unit matrix used to raise the rank of the scalar MFs~\cite{McMullen:1997, Hug:2008}, as
\begin{eqnarray}
    W_0{\mathbf E} &=& W_1^{1,1},\\
    W_1{\mathbf E} &=& W_1^{0,2} + W_2^{1,1}, \\
      W_2{\mathbf E} &=& 2W_2^{0,2}.
    \label{eqn:rank2relation}
\end{eqnarray}
For a single curve Eq.~(5) imply that $W_1^{1,1}$ is proportional to the identity matrix and do not provide any additional information over the scalar MFs. This will also be true for a spatial distribution of non-overlapping curves. Similarly, Eq.~(\ref{eqn:rank2relation}) imply that $W_2^{0,2}$ is proportional to the identity matrix. We will elaborate on how Eq.~(\ref{eqn:rank2relation}) manifests for level sets of random fields in Section~\ref{sec:section3} and argue why it is still useful to analyze it. 
However, $W_2^{1,1}$ does contain very useful additional information in comparison to $W_1$, as explained in the following subsections.

\subsubsection{Intrinsic anisotropy of structures}

The matrix $W_2^{1,1}$ carries information on the shape and alignment
of structures~\cite{Schroder2D:2009}. For a single closed curve the
eigenvalues $\lambda_1$ and $\lambda_2$ of $W_2^{1,1}$ can be shown to
be positive (see Eq. (11) and Appendix C for the proof) and hence chosen such that $\lambda_1 \le
\lambda_2$.  The eigenvalues are also invariant under rotations (they
are completely determined by the trace and determinant). The intrinsic
anisotropy parameter $\beta$ for the curve is then defined as

%Since $W_2^{1,1}$ is a symmetric matrix $\lambda_1$ and $\lambda_2$ are real numbers. They are also positive (for the proof see Eq. (\ref{eqn:positive_eigen})). They are invariant under rotations.

\begin{equation}
\beta\equiv  \frac{\lambda_1}{\lambda_2}.
\label{eq:beta}
\end{equation}

Note that $\beta$ is invariant under scaling the size of the curve.
For some simple anisotropic shapes such as the ellipse it is possible
to derive analytic expressions for $W_2^{1,1}$ and obtain $\beta$. The
expression for an ellipse is given by Eqs. (\ref{eqn:W211formula1})
and (\ref{eqn:W211formula2}) in Appendix B. We refer the reader to
Table 1 of~\cite{Vidhya:2016} for $\beta$ values corresponding to
different aspect ratios of ellipses. See also Fig. 8
of~\cite{Schroder2D:2009} for the case of rectangles.

It is easy to show that any curve that has $m$-fold symmetry with $m\ge 3$ will have $\beta=1$. 
Examples of convex structures that are isotropic are circles, equilateral triangle, square or any equi-angular $n$-polygon\footnote{Normal vectors are ill-defined at the vertices for shapes such as polygons. We can consider the vertices of the polygon to be smoothed. Or we can use the formulae for $W_2^{1,1}$ in pixelized space as done in~\cite{Schroder2D:2009,Vidhya:2016}}.  The isotropy of these shapes is because their main axes are equivalent, even though not all directions are equivalent (except for circle). In general there can be non-convex shapes for which $\beta=1$. {\em A given curve is defined to have isotropic shape if $\beta=1$, and anisotropic if $0<\beta<1$.} The degree of anisotropy is quantified by the departure of $\beta$ from one. For anisotropic shape the eigenvectors corresponding to the eigenvalues pick out two distinct directions that are orthogonal to each other.

%For arbitrary choice of coordinates the eigenvectors will not be aligned with the axes of the coordinates. However, rotation invariance of the eigenvalues implies that $\beta$ will be independent of the orientation of the coordinate axes. This is just a re-statement of what we intuitively understand - the shape of a curve on a plane remains the same regardless of how we choose coordinates.

\subsubsection{Relative alignment of many structures}

\begin{figure}
\begin{center}
  \resizebox{2.4in}{1.45in}{\includegraphics{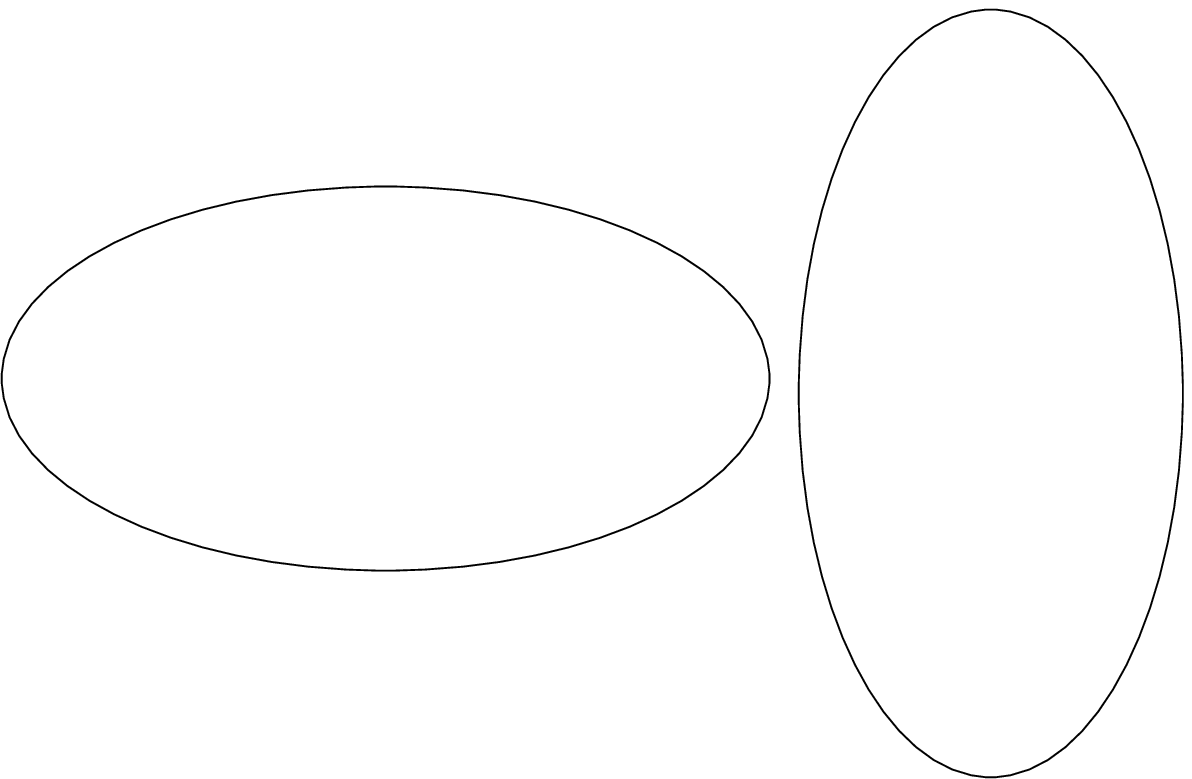}} \hskip 1cm
  \resizebox{.8in}{.8in}{\includegraphics{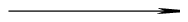}}\hskip 1cm
  \resizebox{1.5in}{1.47in}{\includegraphics{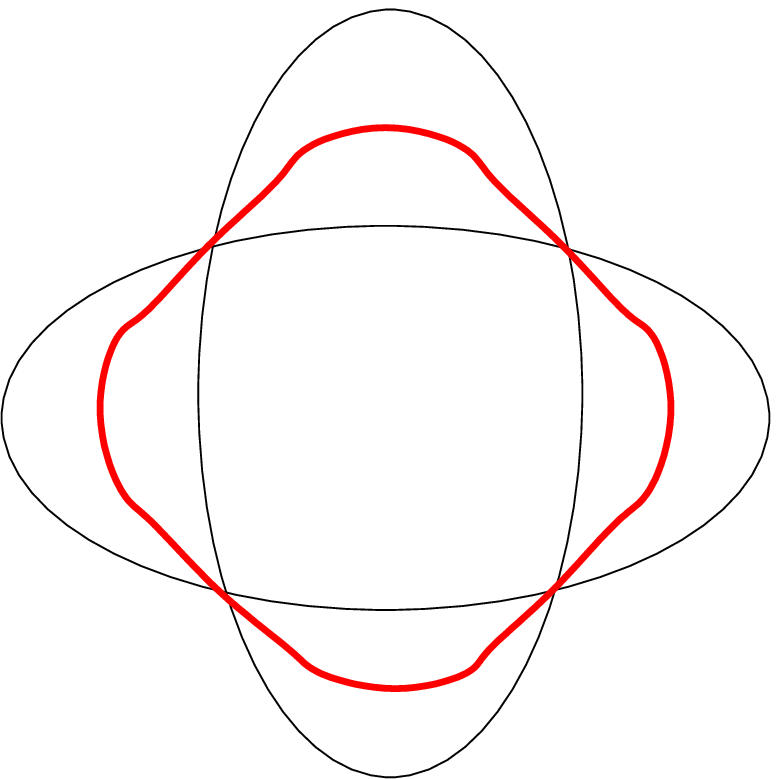}}\\
  \vskip 1cm
  \resizebox{2.85in}{2.4in}{\includegraphics{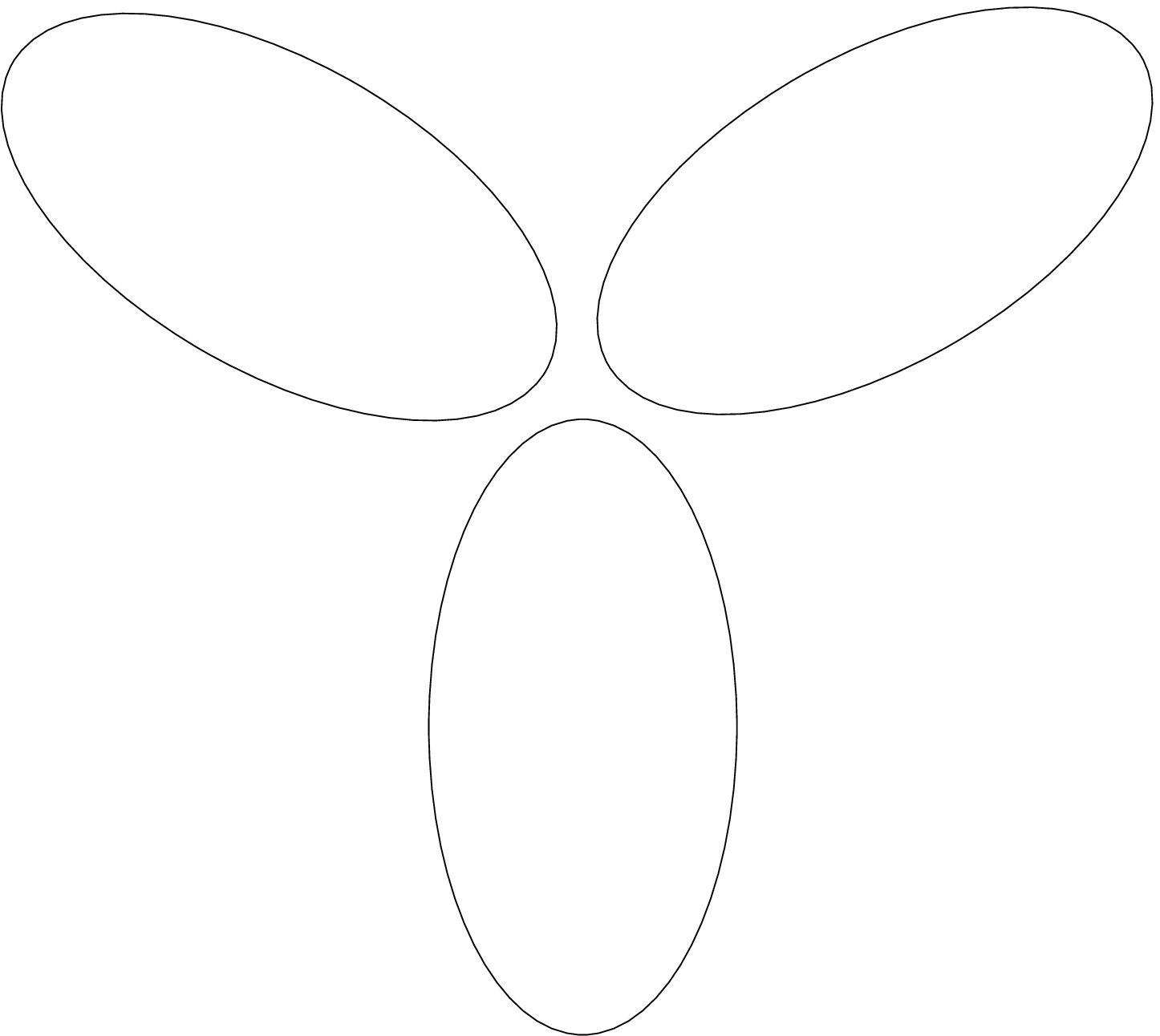}} \hskip .8cm
  \resizebox{.8in}{.8in}{\includegraphics{arrow.eps}}\hskip .8cm
  \resizebox{1.4in}{1.45in}{\includegraphics{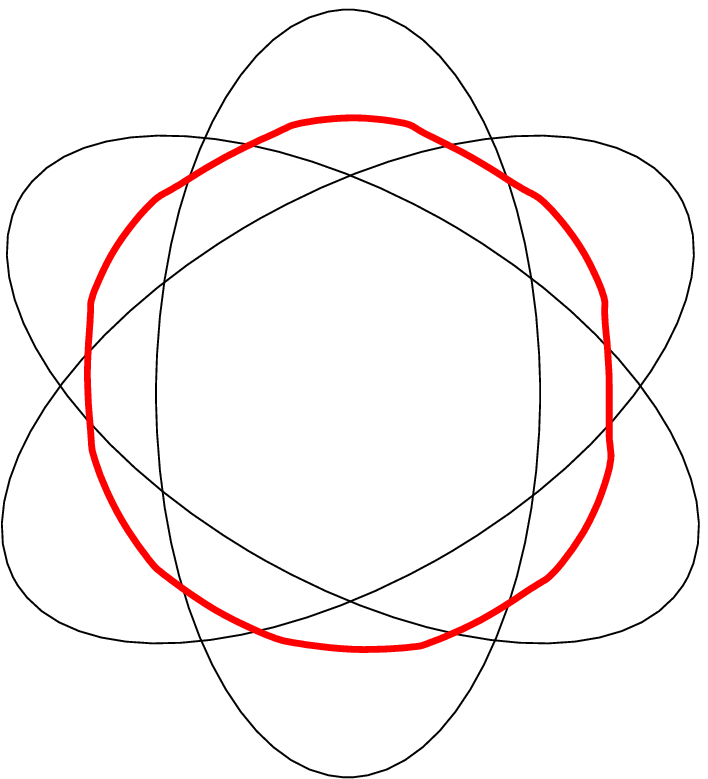}}\\
\end{center}
\label{fig:ellipse_alignment}
\caption{{\em Top row:} Left panel shows two identical ellipses placed with their semi major axes perpendicular to each other. The right panel shows them translated and placed with their centroids overlapping. The red curve shows the locus of points that are the averages of the radial distances of the two ellipses. The locus curve is clearly isotropic. Note that if the ellipses have aspect ratio close to one the locus curve will be convex, whereas, if the aspect ratio is much less than one the locus curve will be non-convex.  {\em Bottom row:} Same as top row but for 3 ellipses placed $120^{\circ}$ to each other. The locus curve is isotropic and looks almost circular.}
\end{figure}

Next consider the case of many closed curves.
Let ${\overline{W}}_2^{1,1}$, be the element by element average of $W_2^{1,1}$ over the set of curves.  
Let $\Lambda_1$ and $\Lambda_2$, such that $\Lambda_1\le \Lambda_2$, denote the eigenvalues of  ${\overline{W}}_2^{1,1}$ (again, the eigenvalues can be shown to be positive as done in Appendix C). Then, we define the ratio $\alpha$ as 
\begin{equation}
\alpha\equiv   \frac{\Lambda_1}{\Lambda_2}.
\label{eq:alpha}
\end{equation}
For a single curve, we recover $\alpha=\beta$. For any distribution of $n$ arbitrary sized circles it is trivial to show that $\alpha=1$.

$\alpha$ encodes the relative alignment between individually anisotropic curves.  To understand this let us first consider two identical ellipses. Let $\delta$ denote the relative angle between their semi major axes. Then using Eq. (\ref{eqn:W211formula1}) and its rotation we can show that if  $\delta=90^{\circ}$ then ${\overline{W}}_2^{1,1}$ is proportional to the identity matrix and hence $\alpha=1$. We get $\alpha < 1$ if $\delta<90^{\circ}$, and $\alpha=\beta$ if $\delta=0$. (See Table 2 of~\cite{Vidhya:2016} for $\alpha$ values corresponding to different relative angles between two ellipses.)

\begin{figure}
\begin{center}
  \resizebox{3.4in}{3.45in}{\includegraphics{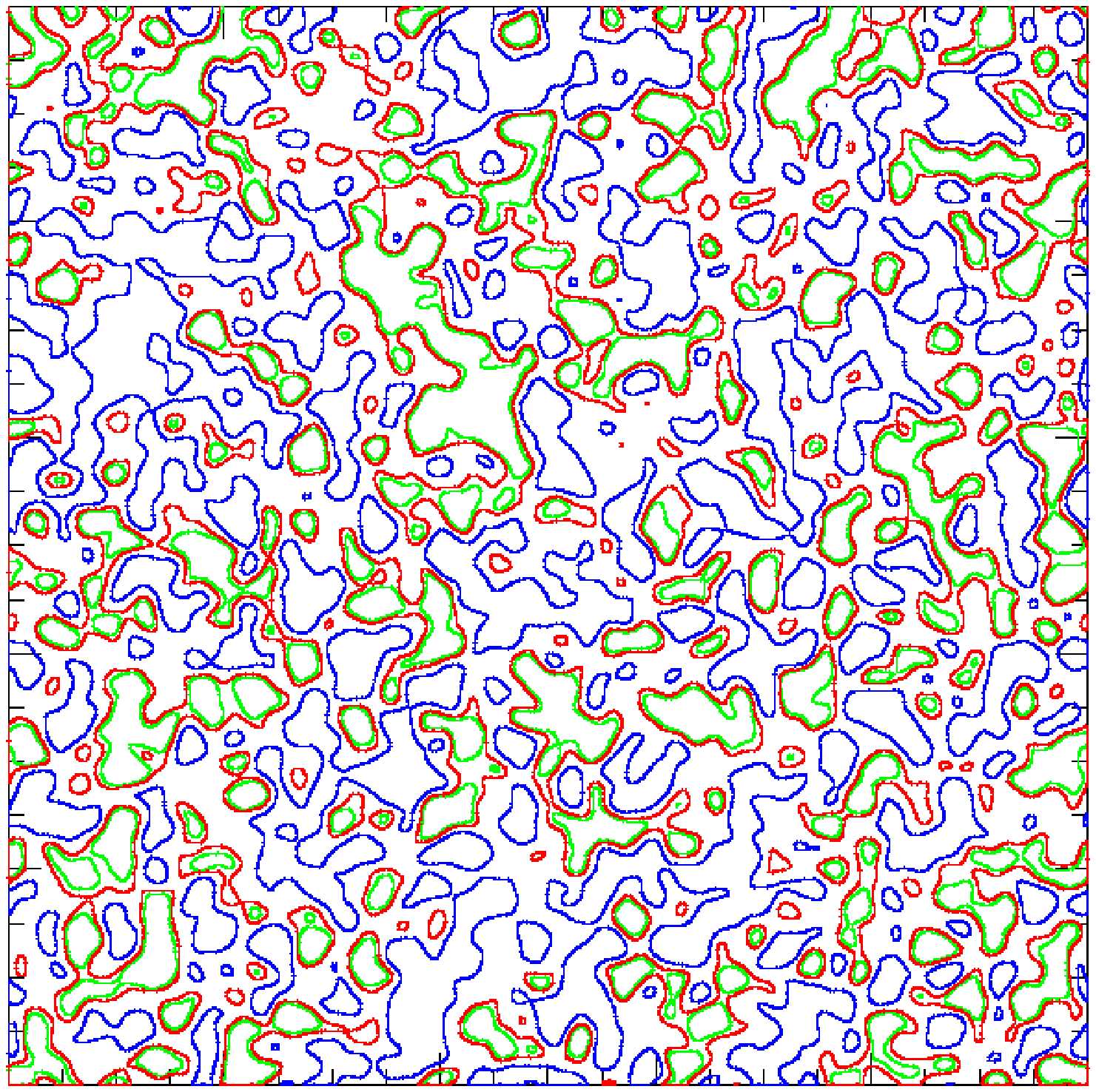}}
\end{center}
\label{fig:random_curves}
\caption{A random spatial distribution of random curves obtained as boundaries of level sets of a random field having zero mean. %The curves of different colours, blue, red, and green, correspond to boundaries of different level sets of the field.
  The  blue curves are boundaries for the level set of a negative threshold field value, red ones are for zero threshold and green for positive.}
\end{figure}

It is instructive to visualize the ellipses as follows. Translation invariance implies that we can translate one of the ellipses such that the two have a common centroid. We refer to the locus of the averages of the radial distances of the ellipses from the center in different directions as the {\em locus curve}. If the locus curve is isotropic (as defined above for a single curve by $\beta=1$) then we expect to get $\alpha=1$.  For example, the top row left panel of Fig. (1) shows two identical ellipses with $\delta=90^{\circ}$. On the right panel the ellipses are translated and superimposed such that their centres overlap. The locus curve is shown in red. It is clear that the locus curve has symmetric shape. If $\delta\ne  90^{\circ}$ the shape of the locus will be anisotropic. 
Note that if the two ellipses are not identical in size we will not get $\alpha=1$ when $\delta=90^{\circ}$. The reason is that the  corresponding locus curve is anisotropic and the level of anisotropy will be determined by the difference in their sizes. For isotropic distribution (equi-angle between semi-major axes) of many identical ellipses the resultant locus would approach a circle as the number of ellipses increases, as is evident from the bottom row panels of Fig. (1) showing 3 ellipses arranged isotropically. 

Hence $\alpha$ gives a measure of the relative alignment or the deviation from rotational symmetry in the spatial arrangement of curves. {\em It is actually the $\beta$ value of the locus curve}. Note that in determining the locus curve the important geometric concepts used are: (1) common centroid, and (2) parallel transport of the curves\footnote{What we mean by parallel transport of the curve is actually parallel transport of the tangent vector to the curve at each point.} along geodesics, which in this case is simply translation. 

 We {\em define} that a given distribution of arbitrary shaped curves is {\em statistically isotropic} or completely unaligned if $\alpha=1$ and anisotropic if $\alpha < 1$. In practical applications, such as to random fields, often we encounter random spatial distribution of random curves. An example is shown in Fig.~(2) where the curves of different colours (blue, red, and green) correspond to boundaries of different level sets of a random field whose mean value is zero. The measurement of $\alpha$ becomes a very useful tool to search for statistical isotropy in such situations.

%%%%%%%%%%%%%%%%%%%%%%%%%%%%%%%%%%%%%%%%%%%%%%%%%%%%%%%%%%%%%%%%%%
\subsection{Definition of rank 2 Minkowski Functionals on the unit sphere}

In this section, we focus specifically on rank-2 TMFs that are translation invariant and generalize their definition to curved space. We then specialize to the unit sphere. Since, from Eq. (5), $W_1^{1,1}$ does not add anything new to the information already encoded in the scalar area MF, $W_0$, we focus here on $W_2^{1,1}$ and $W_2^{0,2}$.   

It has been shown in~\cite{Santalo:1976, Schmalzing:1998} that curvature integrals such as the scalar MFs are well defined on general smooth manifolds, provided $\kappa$ is the geodesic curvature of the curve. In flat space the third scalar MF, $W_2$, is equal to the Euler characteristic, $\chi$, and hence directly gives the difference of counts of clockwise oriented (hotspots) and anticlockwise oriented (coldspots) curve. On spaces of constant curvature, as a consequence of the generalized Gauss-Bonnet theorem~\cite{Allendoerfer:1943,Chern:1944},  $\chi$ is given by a linear combination of $W_0$ and $W_2$. Hence, on the sphere $W_2$ is not directly related to the counts of closed curves. Though either $\chi$ or $W_2$ can be equivalently used as the third scalar morphological descriptor, it  has been standard practice in CMB analyses to use $W_2$.

The tensor-valued MFs, however, cannot be immediately generalized to curved manifolds because the position vector is not a well-defined notion. What is naturally defined is the tangent vector space at every point of the manifold. A curve picks out a unique tangent vector at each point on the manifold that it passes through. And in two dimensions, the unit vector normal to the curve can be obtained by a $90^{\circ}$ rotation of the tangent vector and belongs to the tangent vector space at that point. This $90^{\circ}$ rotation is uniquely defined because the sphere is orientable. 

Integral geometry for tensor quantities on spaces with affine connections has been introduced and studied in~\cite{Moor:1951,Fabian:1957}. Tensorial integration is defined as an inverse operation of covariant derivation. The meaning of the integration is that the tangent vectors are parallel transported to a fiducial point on the curve and then added in the same tangent vector space. The basic notion of tensorial integration is outlined in~\ref{sec:appen}.

Using this notion, we define the following two quantities for a smooth closed curve on a general smooth manifold,  
\begin{eqnarray}
  {\mathcal W}_1 &\equiv& \frac14\,\int_C \, \hat{T} \otimes \hat{T} \,\ds, \nonumber \\
  {\mathcal W}_2 &\equiv& \frac{1}{2\pi}\,\int_C \, \hat{T} \otimes \hat{T} \,\kappa\,\ds, 
\label{eqn:tmf_def}
\end{eqnarray}
where $\hat T$ is the unit tangent vector at each point of the curve. $ {\mathcal{W}}_2$ is the generalization of $ W_2^{0,2}$, and $ {\mathcal{W}}_1$ is the generalization of $ W_2^{1,1}$ to an arbitrary two dimensional manifold. 
$ {\mathcal{W}}_2$ and $ W_2^{0,2}$ are related by a rotation by $90^{\circ}$ between $\hat n$ and $\hat T$. 
$W_2^{1,1}$ contains $\vec r$ explicitly. We can re-express them in terms of the tangent vector and $\kappa$ by using integration by parts (proof given by Eq. (B.9) in Appendix~B), as,
\begin{eqnarray}
    W_2^{1,1}(C) &=& \frac12 \int_{C} \,\hat{T} \otimes \hat{T} \,\ds.
  \label{eqn:tmf_ge}
\end{eqnarray}
%Thus, $ {\mathcal{W}}_j$, with $j=1,2$ are generalizations of  $W_2^{1,1}$ and $ W_2^{0,2}$, respectively.
The expressions of $ {\mathcal{W}}_j$ in terms of the unit tangent vector makes them directly applicable to curved space. Eq. (\ref{eqn:tmf_def}) is our definition of tensor Minkowski Functionals on curved space. 
Note that the numerical factors before the integrals have been chosen so as to match the usual definitions for scalar MFs, $W_1$ and $W_2$, used for CMB fields. 

It is again straightforward to show that the traces of $ {\mathcal{W}}_j$ give the scalar MFs, $W_j$ (see Eqs. (\ref{eqn:traceproof1}) and (\ref{eqn:traceproof2}) in Appendix B for the proofs),
\begin{eqnarray}
  {\mathbf {Tr}}\left({\mathcal{W}}_1\right) &=& \frac14\,\int_C \, \ds = W_1, \label{eqn:tmf_trace1} \\
  {\mathbf {Tr}}\left({\mathcal{W}}_2\right) &=& \frac{1}{2\pi}\,\int_C \,\kappa\,\ds = W_2. 
\label{eqn:tmf_trace2}
\end{eqnarray}

The definitions in Eqs.~(\ref{eqn:tmf_def}) are general and hold for any smooth manifold with affine connection on it. We now focus on the unit sphere for which the isometry group is the orthogonal group, and the orientation preserving subgroup is the rotation group $SO(3)$. 
Under size scaling of the curve, $C \rightarrow \lambda C$, where $\lambda$ is a scaling factor, $\mathcal{W}_j$, transform as 
  \begin{eqnarray}
   \mathcal{W}_1 (\lambda C) &=& \lambda \mathcal{W}_1(C),\nonumber\\
   \mathcal{W}_2 (\lambda C) &=&  \mathcal{W}_2(C),
\label{eqn:scaling}
 \end{eqnarray}
Eq.~(\ref{eqn:scaling}) follows because under scaling the magnitude of $\hat T$ is invariant, $\kappa$ scales as $\lambda^{-1}$ and ${\rm d}s$  as $\lambda$.

%%%%%%%%%%%%%%%%%%%%%%%%%%%%%%%%%%%%%%%%%%%%%%%%%%%%%%%%%%%%%%%%
\subsubsection{Intrinsic anisotropy and alignment measures for closed curves on the unit sphere}

The intrinsic shape and alignment measures defined in subsections 2.1.1 and 2.1.2 on flat space can be generalized to the unit sphere.  The eigenvalues of $\mathcal{W}_1$, which we again denote by  $\lambda_1$ and $\lambda_2$ such that $\lambda_1 \le \lambda_2$, are invariant under rotations.  
The intrinsic shape of the curve is captured by the parameter
\begin{equation}
\beta\equiv  \frac{\lambda_1}{\lambda_2}.
\label{eqn:beta}
\end{equation}
The curve is defined to have isotropic shape if $\beta=1$. For a circle on the sphere this definition is a trivial extension from the notion on flat space since a circle always lies on a plane that cuts the sphere. 
Curves that have $0<\beta<1$ are defined to be anisotropic.  

For the case of many closed curves, again  
let $\Lambda_1$ and $\Lambda_2$, such that $\Lambda_1\le \Lambda_2$, denote the eigenvalues of  ${\overline{\mathcal W}}_1$, where the overbar denotes averaging for each element of ${\mathcal W}_1$ over the set of curves.  
Then, we define the ratio $\alpha$ as 
\begin{equation}
\alpha\equiv   \frac{\Lambda_1}{\Lambda_2}.
\label{eqn:alpha}
\end{equation}
Generalizing from the case of flat space, we will say that a random distribution of curves is isotropic if we obtain $\alpha=1$, and anisotropic if $\alpha < 1$.

In order to interpret $\alpha$ as  the $\beta$ value of the locus curve, as done for flat space in section 2.1.2, we need to first identify the centroid of the curve. This is given by  the {\em Riemannian centre of mass}, which is defined to be the point that minimizes the sum of the squares of geodesic distances on curved manifolds with affine connection~\cite{Karcher:1973,Karcher:2014}. Secondly, we need to translate the curves along geodesics so that the respective centre of mass points coincide. Even though this geometric construction is intuitively clear, the mathematical details are not obvious. We will present the mathematical details elsewhere as follow up work. For the purpose of this paper we proceed here with $\alpha=1$ as the definition of statistical isotropy for the collection of curves.  

%%%%%%%%%%%%%%%%%%%%%%%%%%%%%%%%%%%%%%%%%%%%%%%%%%%%%%%%%%%%%%%%
\section{TMFs for random fields on the unit sphere, $\mathcal S^2$}
\label{sec:section3}

For a cosmological random field the level set or excursion set associated with each threshold choice of the field provides a distribution of non-intersecting~\footnote{There can be saddle points of the field where two closed curves may meet. Such points can lead to confusion in the counting of curves.} smooth (infinitely differentiable at every point) curves 
(see Fig.~(2)). 
The TMFs associated with these curves will vary systematically as we vary the threshold. For each excursion set the line integrals in Eqs. (\ref{eqn:tmf_def})  can be transformed to area integrals by introducing $\delta-$function and a suitable Jacobian as done in ~\cite{Schmalzing:1998}. This gives
\begin{eqnarray}
    {\overline{\mathcal W}}_1 &=& \frac14 \int_{{\mathcal S}^2} \da \ \delta(u-\nu_t)\ |\nabla u| \ \hat{T} \otimes \hat{T},\\
    {\overline{\mathcal W}}_2 &=& \frac1{2\pi} \int_{{\mathcal S}^2}  \da \ \delta(u-\nu_t)\ \kappa\,|\nabla u| \
    \hat{T} \otimes \hat{T}, 
\end{eqnarray}
where $u$ is the field, $\nu_t$ is the threshold, and $\da$ is the area element. We need to express $\hat T$ in terms of the field. The vector normal to the curve is given by $\vec n= \nabla u = (u_{;1},u_{;2})$, where $\nabla$ denotes the covariant derivative on the sphere and $u_{;i}$ is the $i$-th component of the covariant derivative. Hence we can choose each component of $\hat T$ as
\begin{equation}
  {\hat T}_i = \epsilon_{ij} \frac{u_{;j}}{|\nabla u|},
  \end{equation}
where $\epsilon_{ij}$ is the antisymmetric tensor with $\epsilon_{12}=1$.
$\kappa$ is given by (see Appendix B for proof),% of ~\cite{Schmalzing:1998})
\begin{equation}
  \kappa = \frac{ 2u_{;1}u_{;2}u_{;12} - u_{;1}^2 u_{;22} - u_{;2}^2 u_{;11}}{{|\nabla u|^3}}.
\end{equation}
Then we get
\begin{eqnarray}
  {\overline{\mathcal W}}_1 &=& \frac14\,\int_{{\mathcal S}^2}  \da  \ \delta(u-\nu_t)\ \frac{1}{|\nabla u|} \ {\mathcal M},\\
    {\overline{\mathcal W}}_2&=& \frac1{2\pi}\,\int_{{\mathcal S}^2}  \da \ \delta(u-\nu_t)\ \frac{\kappa}{|\nabla u|} \ {\mathcal M},
\end{eqnarray}
where the matrix $\mathcal M$ is
\begin{equation}
  \mathcal M=  \left(
  \begin{array}{cc} 
    u_{;2}^2 &  u_{;1} \,u_{;2} \\
      u_{;1} \,u_{;2} & u_{;1}^2
  \end{array}\right).
  \label{eqn:M}
  \end{equation}
For any square matrix the determinant and the trace are invariant
under orthogonal transformations. The determinant of $\mathcal M$ is
zero at every point on the sphere. It is easy to check that
${\overline{\mathcal W}}_1$ and ${\overline{\mathcal W}}_2$ will,
however, have non-zero determinant. %(The same is true for ${\mathcalW}_1$ and ${\mathcal W}_2$).
  If $u_{;1}$ and $u_{;2}$ are
uncorrelated the off-diagonal terms of ${\overline{\mathcal W}}_1$ and
${\overline{\mathcal W}}_2$ will always be zero. Then the diagonal 
elements will be the eigenvalues and their product will give the
determinant. The trace of ${\overline{\mathcal W}}_1$ is given by
\begin{eqnarray}
  \sum_i \left( {\overline{\mathcal W}}_1\right)_{ii} &=& \frac14\,\int_{{\mathcal S}^2}  \,\da\, \delta(u-\nu_t)\
  \frac{1}{|\nabla u|} \ \left( u_{;1}^2 + u_{;2}^2\right) \nn\\ 
       {}& =& \frac14\,\int_{{\mathcal S}^2}  \,\da\, \delta(u-\nu_t)\ |\nabla u|,
       \label{eqn:Wtrace}
\end{eqnarray}
which is the second scalar MF - the contour length. This simply reproduces Eq.~(\ref{eqn:tmf_trace1}) and proves them for random fields.  
The trace of ${\overline{\mathcal W}}_2$ is given by
\begin{eqnarray}
  \sum_i \left( {\overline{\mathcal W}}_2\right)_{ii} &&= \frac1{2\pi}\,\int \,\da\, \delta(u-\nu_t)\
  \frac{\kappa}{|\nabla u|} \ \left( u_{;1}^2 + u_{;2}^2\right) \nn\\
       {}&=& \frac1{2\pi}\,\int \,\da\, \delta(u-\nu_t)\
  {\kappa}\,{|\nabla u|} \nn\\
{} &=&        \frac1{2\pi}\,\int \,\kappa\, \ds,
\end{eqnarray}
which is the third scalar MF - the genus.  This reproduces Eq.~(\ref{eqn:tmf_trace2}).

For practical applications the space is pixellized and we need to compute the TMFs numerically. To do so, the $\delta$ function can be approximated as~\cite{Schmalzing:1998},
\begin{equation}
\delta(u-\nu_t) = \frac{1}{\Delta \nu_t}
\end{equation}
when $u$ lies between $\nu_t-\Delta\nu_t/2$ to $\nu_t+\Delta\nu_t/2$, and zero otherwise. Using this, and incorporating masking of parts of the sky, we can express the TMFs {\em per unit area }, denoted by $\overline{w}_j$, as,
\begin{equation}
{\overline{w}}_j \ = \ \frac{\sum_k \omega(k)\,\mathcal{I}_j(k)}{\sum_k\omega(k)},
\label{eqn:wi}
\end{equation}
where $j=1,2$, each pixel is indexed by $k$ and $\omega(k)$ is one if the pixel lies inside the unmasked region, and zero if masked. 
The functions $\mathcal{I}_i$ are given by 
\begin{eqnarray}  
  \mathcal{I}_1(k) &=& \frac{1}{\Delta \nu_t} \,\frac1{|\nabla u|}\, {\mathcal M},  \label{eqn:I1} \\ 
  \mathcal{I}_2(k) &=& \frac{1}{\Delta \nu_t}\, \frac{\kappa}{|\nabla u|} \,{\mathcal M},
  \label{eqn:I2}
\end{eqnarray}
where the RHS is to to be computed at each pixel $k$.

Note that the formalism described here holds for any curved space where the covariant derivative is well defined. It reduces to the case of flat space by simply replacing the covariant derivatives to the usual derivatives. Hence Eqs. (25), (26) and (27) can be used to compute average TMFs for fields on flat two dimensional space.
%%%%%%%%%%%%%%%%%%%%%%%%%%%%%%%%%%%%%%%%%%%%%%%%%%%%%%%%%%%%%%%%%%%%%%%%%
\subsection{Ensemble expectation values for isotropic Gaussian fields}
\label{sec:gaussian}

Next to get the ensemble expectation value for isotropic Gaussian field $u$, we can take the joint Gaussian PDF of $u$, $u_{;i}$ and $u_{;ij}$ and integrate. Assuming isotropy, let $\xi(r)$ be the correlation function of $u$ where $r$ denotes the distance between two spatial points. Then the variances of $u$, $u_{;i}$ and $u_{;ij}$ are given by 
\begin{eqnarray}
  \sigma_0^2 &=& \langle u^2\rangle = \xi(0)\nonumber\\
  \sigma_1^2 &=& \langle |\nabla u|^2\rangle =  2\xi''(0) \nonumber\\
  \sigma_2^2 &=& \langle |\nabla^2u|^2\rangle = 2\xi''''(0) 
\end{eqnarray} 
Consider the 6 component vector $\vec{X}=u,\,u_{;1},\,u_{;2},\,u_{;11},\,u_{;22},\,u_{;12}$. The joint probability distribution of $\vec X$ is given by the form
\begin{equation}
  P(\vec{X}) = \frac1{\sqrt{(2\pi)^6 \,{\rm det} \Sigma }} \,\exp\left(-\frac12 \,\vec{X}\,{\Sigma}^{-1}\,\vec{X} \right), 
\label{eqn:gpdf}
\end{equation}
where the covariance matrix $\Sigma$~\cite{Tomita:1986} is given by
\begin{equation}
  \Sigma = \left(
  \begin{array}{cccccc}
    \sigma_0 &0&0&- \sigma_1/2 &-\sigma_1/2 &0\\
    0& \sigma_1/2 &0&0&0&0\\
    0&0& \sigma_1/2 &0&0&0\\
    -\sigma_1/2 & 0&0& \sigma_2/2 & \sigma_2/6&0\\
    -\sigma_1/2 &0&0& \sigma_2/6& \sigma_2/2 &0\\
    0&0&0&0&0& \sigma_2/6
  \end{array}
  \right)
\end{equation}
Then, the ensemble expectation value for each $\nu$, is
\begin{eqnarray}
    \big\langle {\overline{\mathcal W}}_1(\nu_t) \big\rangle &=& \int {\rm d}\vec{X}\, P(\vec{X})\, \int_{{\mathcal S}^2}  \da\, \delta(u-\nu_t)\, \frac{1}{|\nabla u|}  \, {\mathcal M} \nonumber \\
    \big\langle {\overline{\mathcal W}}_2(\nu_t) \big\rangle &=& \int {\rm d}\vec{X}\, P(\vec{X})\, \int_{{\mathcal S}^2}  \da\,\delta(u-\nu_t)\, \frac{\kappa}{|\nabla u|}   \, {\mathcal M}. \nonumber\\
    &{}& 
    \label{eqn:ensemblex}
\end{eqnarray}
If the field is isotropic, the area and $\vec X$ integrations must commute. So we can carry out the $\vec X$ integration first. This will not be the case if the field is not isotropic. 
Defining the correlation length, $r_c\equiv \sigma_0/\sigma_1$ and the normalized threshold $\nu\equiv\nu_t/\sigma_0$, we get the ensemble expectation per unit area to be
\begin{eqnarray}
  \big\langle {\overline{w}}_1(\nu) \big\rangle&=&
  \frac{1}{16\sqrt{2}\ r_c}\, e^{-\nu^2/2} \ \id,\\
   \big\langle {\overline{w}}_2(\nu) \big\rangle&=&
   \frac{1}{8\sqrt{2} \,\pi^{3/2}\ r_c^2} \ \nu\,e^{-\nu^2/2} \ \id,
   \end{eqnarray}
where $\id$ is the identity matrix. Note that this result holds even if the space has infinite extent. 
The alignment parameter is obtained to be $\alpha=\langle {\overline{w}}_1\rangle_{11}/ \langle {\overline{w}}_1\rangle_{22} =1$ at every $\nu$.  

\vskip .4cm
The information encoded in the $\big\langle\overline{w}_1\big\rangle$ can be summarized as follows,
\begin{itemize}                                                                 
\item {\em Off-diagonal elements}: For a Gaussian field, regardless of any departure from isotropy, $u_{;1}$ and $u_{;2}$ are always uncorrelated. Therefore, the off-diagonal elements of ${\overline{\mathcal W}}_1$ are always zero. The same argument holds for  ${\overline{\mathcal W}}_2$ also.
\item {\em Gaussian nature}: The Gaussian nature of the field is encoded in the specific functional form of the diagonal elements. Their sum gives the expectation value for the scalar MF $W_1$.                   
\item {\em Cosmological parameters}:  are  encoded in the correlation length, $r_c$.                            
\item {\em Statistical isotropy}:  is encoded in the identity matrix, or in the equality of the diagonal matrix elements. We recover $\alpha=1$. 
As explained in sections 2.1.2 and 2.2.1, what this means is that the ensemble average of structures in fluctuations, such as hot spots or coldspots in 2-dimensions, must be circular and depend on the threshold. The effective radius of the circle must be given by $\frac{1}{r_c} e^{-\nu^2/2}$ upto a numerical factor. 
\end{itemize}

$\big\langle\overline{w}_2\big\rangle$ will, however, always be proportional to the identity matrix, regardless of the isotropy/anisotropy of the field. The two diagonal elements can provide independent measurements of any departure of the field from Gaussian nature. This extra information can help tighten constraints in non-Gaussian searches.

It is useful to note that the analytic expressions obtained here are general and hold in flat as well as curved spaces.

%%%%%%%%%%%%%%%%%%%%%%%%%%%%%%%%%%%%%%%%%%%%%%%%%%%%%%%%%%%%%%%%%%%
\subsection{Ensemble expectation values for isotropic CMB polarization intensity - Rayleigh fields}

The total polarization intensity of the CMB is given by
\begin{equation}
I_{\rm Pol}=\sqrt{Q^2+U^2}, 
\end{equation}
where $Q$ and $U$ are the usual Stokes parameters. For $Q$ and $U$ sourced by primordial density perturbations which are Gaussian in nature, they are Gaussian with zero mean and same variance, and the PDF of $I_{\rm pol}$ will have Rayleigh form. It was pointed out in~\cite{Chingangbam:2017} that the equality between the variances of $Q$ and its derivatives and the corresponding variances of $U$ break down for partial sky due to the spin-2 nature of $Q$ and $U$. Here, we consider only full sky coverage. 

Analytic expressions for the scalar MFs for  $I_{\rm pol}$ were derived in~\cite{Naselsky:1998by}. Here we extend their calculation to TMFs. Consider the 12 dimensional vector $\vec X \equiv (Q,\, U, Q_{;i}, U_{;i},\, Q_{;ij}\, U_{;ij})$, where $i,j=1,2$. The joint PDF of $\vec X$ is given by
\begin{equation}
  P(\vec{X}) = \frac1{\sqrt{(2\pi)^{12} \,{\rm det} \Sigma }} \,\exp\left(-\frac12 \,\vec{X}\,{\Sigma}^{-1}\,\vec{X} \right),
\label{eqn:ppdf}
\end{equation}
where $\Sigma$ is the covariance matrix (see section 2.2 of~\cite{Naselsky:1998by} for the expression). 
Then, by following similar calculation as done in Section~\ref{sec:gaussian} we obtain
\begin{eqnarray}
  \big\langle \overline{w}_1(\nu) \big\rangle&=&
  \frac{1}{{2}\ r_c} \nu \,e^{-\nu^2/2} \ \id,\\
     \big\langle {\overline{w}}_2(\nu) \big\rangle&=&
     \frac{1}{8\pi\ r_c^2} \ (\nu^2-1)\,e^{-\nu^2/2} \ \id,
     \label{eqn:Ptmf}
\end{eqnarray}
where $\nu$ and $r_c$ are as defined in the previous subsection. 
The interpretation of the information encoded in Eq. (\ref{eqn:Ptmf}) is the same as the case of the Gaussian isotropic field.

%%%%%%%%%%%%%%%%%%%%%%%%%%%%%%%%%%%%%%%%%%%%%%%%%%%%%%%%%%%%%%%%%%%%%%%%%%%%%
\section{Some applications}

In order to compute the TMFs numerically for any given field we can use Eqs. (\ref{eqn:wi}),  (\ref{eqn:I1}) and  (\ref{eqn:I2}). In this section we first present calculations of $\alpha$ for simulated CMB temperature maps to show that we get the expected statistically isotropic behaviour at different threshold values of the field. Then we apply to maps of foreground emissions provided in the 2015 PLANCK data release.

\subsection{$\alpha$ for simulated temperature maps}

We first test the computation on simulated Gaussian and isotropic CMB temperature maps. The input $C_{\ell}$ was obtained using {\texttt{CAMB}}~\cite{Lewis:2000ah,cambsite} and the input best fit $\Lambda$CDM cosmological parameter values taken from~\cite{planck:cosmopara2015}.
The maps are then simulated using {\texttt{HEALPIX}}~\cite{Gorski:2005,Healpix}.\footnote{http://healpix.sourceforge.net}.
\begin{figure}
\begin{center}
\resizebox{4.in}{3.2in}{\includegraphics{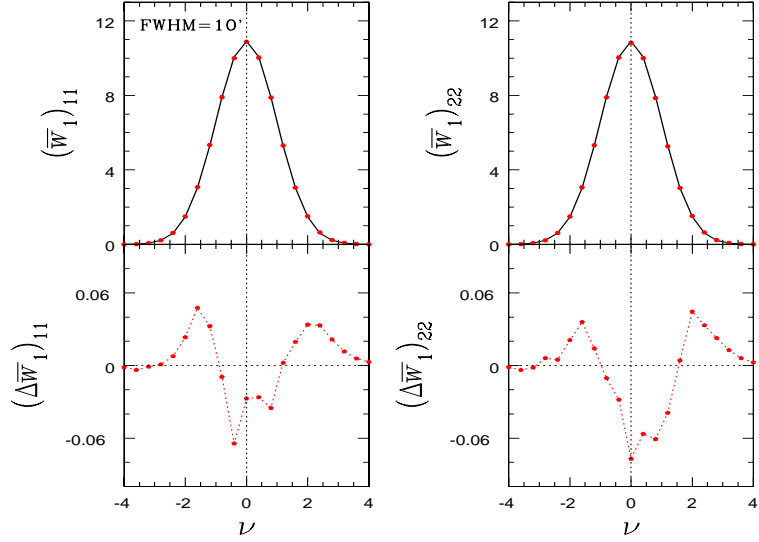}}\quad
\end{center}
\label{fig:w1_gaussian}
\caption{Upper panels are plots of diagonal elements of $\mathcal W_1$ obtained from one simulated Gaussian isotropic CMB temperature map. Black line is Gaussian analytic expected given by Eq. (31), where $\sigma_0$ and $\sigma_1$ have been calculated using the simulated map. Red dots are results of numerical computation using Eqs. (25) from the same temperature map.  The lower panels are the numerical error obtained by subtracting the numerical results from the analytic expectation. This error agrees well with the error that is expected due to the discretization of $\delta$ function given by Eq. (\ref{eqn:dw1_g}). }
\end{figure}

The numerical errors due to the $\delta-$function approximation in the calculation of scalar MFs for a Gaussian field was estimated in~\cite{Lim:2012}. For the diagonal elements of ${\overline{w}}_i$ the error can be quantified as
\begin{equation}
{\overline{w}}_j \ = \ {{\overline{w}}_j}^{\rm {G,ana}} + \Delta {\overline{w}}_j,
\label{eqn:dw1}
\end{equation}
where $j=1,2$, the superscript (G,ana) stands for Gaussian analytic formula. For ${\overline{w}}_1$,  $\Delta {\overline{w}}_1$ is given by (see Eq. (3.9a) of ~\cite{Lim:2012})
\begin{equation}
  \Delta {\overline{w}}_1 = 
  \frac{\sqrt{\pi}}{4\sqrt{2}}\ \frac{\sqrt{\tau}}{\Delta \nu }\
  \bigg( {\rm erf}(\nu_+) - {\rm erf}(\nu_-) \bigg)  
 - {{\overline{w}}_1}^{\rm{G,ana}} (\nu)
\label{eqn:dw1_g}
\end{equation}
where $\nu_+=\nu+\Delta\nu/2$ and $\nu_-=\nu-\Delta\nu/2$.
\begin{figure}
\begin{center}
\resizebox{2.2in}{2.2in}{\includegraphics{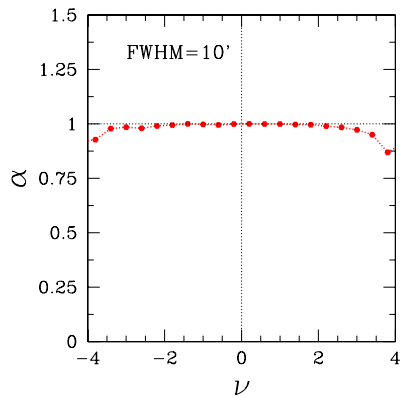}}
\end{center}
\label{fig:alpha_gaussian_iso}
\caption{$\alpha$ versus $\nu$ for a Gaussian isotropic temperature map. $\alpha=1$ is recovered very well at threshold values close to 0 where the number of structures is the largest. At higher $|\nu|$ the values of $\alpha$ deviate from one because the number of structures become fewer and for a random distribution of a few structures the probability that they will be isotropically distributed is very small.}
%due to a combination of two reasons - statistical fluctuation caused by fewer number of structures and the ordering of the eigenvalues.}
\end{figure}
The top panels of Fig.~(3) show the diagonal elements of $\overline{w}_1$ for one isotropic Gaussian CMB temperature map. The bottom panels show the numerical error given by Eq.~(\ref{eqn:dw1}) and it agrees well with the analytic expectation given by Eq.~(\ref{eqn:dw1_g}). The bin size used is $\Delta\nu=0.4$. The resulting values of $\alpha$ as a function of the threshold is shown in Fig.~(4). 
$\alpha=1$ is recovered very well at threshold values close to the mean temperature value zero where the number of structures is the highest and the total length of the curves is the largest. At higher $|\nu|$ the values of $\alpha$ deviate away from one. % caused by a combination of two factors - statistical fluctuation caused by fewer number of structures, and the ordering of the eigenvalues of $\overline{w}_1$.
The reason for this is that as $|\nu|$ increases there are fewer structures. For a random distribution of a few structures the probability that they will be arranged isotropically is very small. As the number of structures decreases further $\alpha$ will tend to the value of $\beta$ of the last structure.

%%%%%%%%%%%%%%%%%%%%%%%%%%%%%%%%%%%%%%%%%%%%%
\subsection{Application to Galactic foreground fields}

Understanding the statistical properties of the foreground fields are valuable for understanding how they contaminate the true CMB fields, apart from of course their own intrinsic astrophysical importance. 
Non-Gaussianity of PLANCK foreground maps have been analyzed in~\cite{Ben-David:2015}. As mentioned in the introduction, ~\cite{Chingangbam:2012} used scalar MFs to detect the presence of residual Galactic foreground and point sources contamination in the WMAP data. 
Here, in order to demonstrate the measure of alignment and departure from Gaussianity using $\mathcal{W}_1$ we apply the method to maps of the Galactic foreground temperature emissions provided as part of the 2015 PLANCK data release%\footnote{Based on observations obtained with Planck, an ESA science mission with instruments and contributions directly funded by ESA Member States, NASA, and Canada.}
~\cite{Adam:2015,PLA}. 
Our intention here is not that of carrying out a detailed analysis of the morphology and non-Gaussianity of the foreground maps, but rather to demonstrate how $\alpha$ informs us about the statistical anisotropy and non-Gaussianity of some example fields.

 The separation of diffuse foreground components has been carried out by the PLANCK team on a combination of the PLANCK observations with the 9 year WMAP sky maps~\cite{WMAP9:2013} and the Haslam et al~\cite{Haslam:1982} 408 MHz map.
Bayesian analysis is used to fit the models for the various diffuse foreground components to the observed data by using the so-called \texttt{COMMANDER} code. 
We refer to Table 4 of~\cite{Adam:2015} for a summary of the models which are typically characterized by model parameters such as the spectral index, reference frequency, and the intensity for the reference frequency. The chosen priors for model parameters of the different foreground components are also given in this table. We also refer to Table 5 of the same reference which lists the component separated maps with the best fit model parameters. The maps we use here are temperature maps of synchrotron, thermal dust, anomalous microwave emissions (AME) and the CO, $j=1 \rightarrow 0$ line emissions. For completeness we describe each map briefly.

\begin{figure*}
\begin{center}
  \resizebox{6.in}{4.7in}{\includegraphics{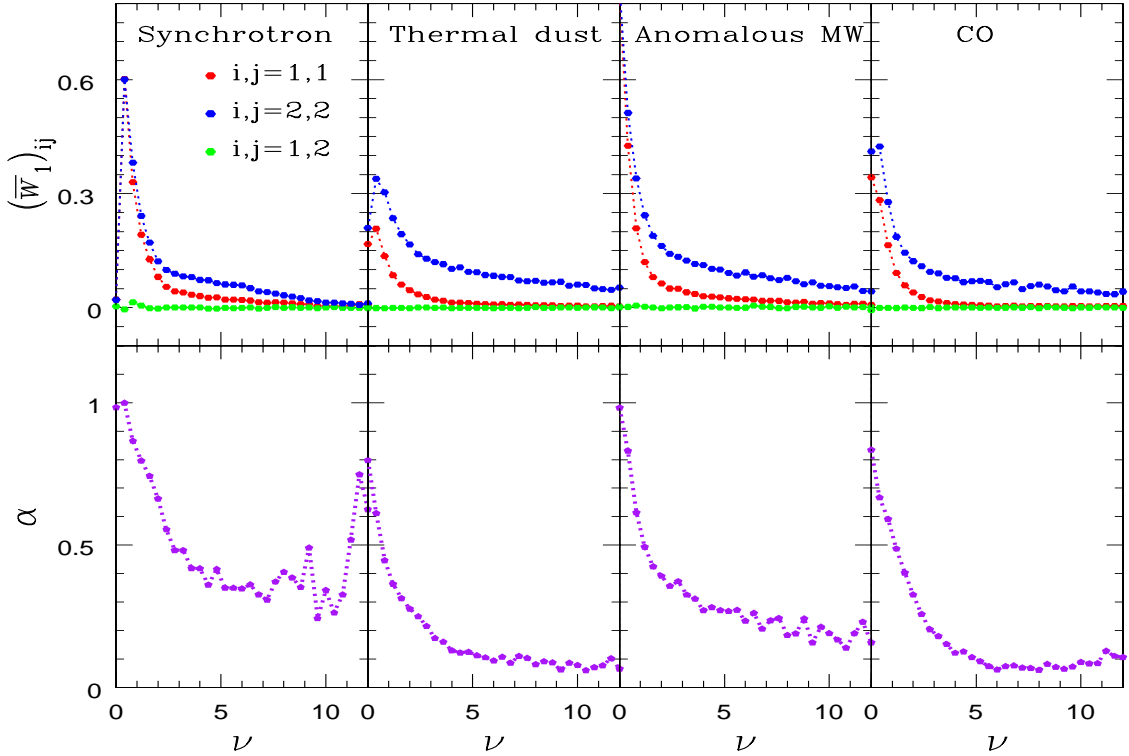}}
  \label{fig:w1_planckfg}
\end{center}
\caption{The top panels are plots of the elements of $\overline{w}_1$ for simulated synchrotron, thermal dust, anomalous microwave and CO ($J=1\rightarrow 0$) emissions of our Galaxy taken from the 2015 PLANCK data release. The bottom panels show the corresponding $\alpha$ values for these fields.}
\end{figure*}

For synchrotron emission the brightness temperature at each pixel is given by
\begin{equation}
  S_s(\nu)=A_s \left(\frac{\nu_0}{\nu}\right)^2\frac{f_s(\nu/\alpha)}{f_s(\nu_0/\alpha)},
\end{equation}
where $\nu_0=408 \ {\rm MHz}$, and  $f_s(\nu/\alpha)$ is a template given by the \texttt{GALPROP} model (see~\cite{Orlando:2013} and references therein). The best fit value of the parameter $\alpha$ (not to be confused with the alignment parameter) is 0.26. 
For dust emissions the brightness temperature at each pixel is given by
\begin{equation}
  S_{\rm d}(\nu)=A_{\rm d}\left(\frac{\nu}{\nu_0}\right)^{\beta_d+1}\frac{\exp(\gamma\nu_0)-1}{\exp(\gamma\nu)-1},\quad \gamma=h/(k_BT_d),
\end{equation}
where $\nu_0=545 \ {\rm GHz}$, mean value of $\beta_d$ is 1.55 and the grey body temperature $T_d$ has a mean value of 23 K. 

The AME (spinning dust) is modelled using two dust components, with the brightness temperature of each component given by
\begin{equation}
  S_{\rm sd}^{i}(\nu)=A_{\rm sd}^{i}\left(\frac{\nu_0}{\nu}\right)^2\frac{f_{\rm sd}(\nu\nu_{p0}/\nu_p^{i})}{f_{\rm sd}(\nu_0^i\nu_{p0}/\nu_p^{i})},
\end{equation}
where the subscript $i$ takes values 1 or 2 for the two components. $\nu_0^1=22.8 \ {\rm GHz}$, $\nu_0^2=41 \ {\rm GHz}$, $\nu_{p0}=30\ {\rm GHz}$. The first spinning dust component is modelled with a free peak frequency $\nu_{p}^1$ with mean value $19\ {\rm GHz}$, while the second component has  spatially constant peak frequency $\nu_{p}^2$.   
$f_{\rm sd}(\nu)$ is given by an external template function.  
The CO line emission map we use is the $J=1\rightarrow 0$, so-called Type 1 map. This emission is also similarly characterized by an amplitude, $A_{\rm CO}$. The amplitudes of the brightness temperatures of the various foreground components, namely, $A_s, A_d, A_{\rm sd}^1, A_{\rm sd}^2, A_{\rm CO}$, along with the amplitudes of other components that we have not considered here, are then determined pixel by pixel by carrying out multi-component, multi-frequency Bayesian fitting to observed data from PLANCK, 9 year WMAP data and the Haslam et al. 408 MHz map.

The resolution of the maps we use here correspond to $1^{\circ}$ FWHM with the \texttt{HEALPIX} parameter value Nside$=256$. 
We work with the fields rescaled by their respective rms values. Note that no masking of the maps have been done, which means the main contributions to the results that we present below come from the Galactic region where most of the foreground emissions are localized. To interpret our results we use the expected Gaussian shape of the elements of $\overline{w}_1$ and $\alpha$ values given for the simulated Gaussian and isotropic CMB temperature map given in the above subsection 4.1 as reference. 
The top panels of Fig.~(5) show the elements of $\overline{w}_1$ for synchrotron, thermal dust, anomalous microwave and CO emissions. The threshold values on the $x$-axis are all positive since these are emissions and we have not subtracted the mean, unlike the CMB maps which are fluctuations about the mean. By visual inspection, we can see that both the diagonal elements of $\overline{w}_1$ for all the fields are very different from Gaussian functions (compare with the expected functional form for Gaussian field given by Eq. (34) or Fig. (3)). Hence they are highly non-Gaussian. This is as expected since foreground emissions are not Gaussian fields. Further, the nature of non-Gaussianity for the different fields are different. 

The lower panels of Fig.~(5) show the corresponding $\alpha$ for each of the foreground  maps. We find that all maps show high level of alignment at most of the threshold values. Therefore, the fields are statistically highly anisotropic. This corroborates with what can be expected from visual inspection of the foreground maps where we can see directional patterns along the Galactic plane. This implies that there is coherence in the distribution of the matter emitting these emissions. Further, the level of alignment differs between the different emissions. This indicates that there are variations in the spatial distribution of emitting materials. Maps of thermal dust and CO show higher level of alignment in comparison to synchrotron and AME.

%%%%%%%%%%%%%%%%%%%%%%%%%%%%%%%%%%%%%%%%%%%%%%%%%%%%%%%%%%%%%%%%%%%%%%%%%%

\section{Conclusion}

% ADD PARA about extra info
In order to clarify the new information that can be obtained from the translation invariant rank-2 TMFs in comparison to the scalar MFs, we  reiterate that there are 3 independent tensors, each of which is a tensor generalization of the corresponding scalar MFs. The tensor generalization, $W_1^{1,1}$, of the area fraction $W_0$ does not yield any new information, and hence we have not focused on it in this paper. The tensor generalization of the contour length,  $W_2^{1,1}$, provide new information about the anisotropy of the shape of structures and the relative alignment between many structures. In this paper we have provided new geometric insight into the intrinsic anisotropy and statistical isotropy of structures encoded in the TMFs in flat 2-dimensional space. When we compute the trace to obtain the scalar contour length from $W_2^{1,1}$, we are reducing from two degrees of freedom to one. In doing so we are effectively throwing away the anisotropy and alignment information.

To the best of our knowledge TMFs for random fields have not been studied in the mathematics literature, neither for flat nor curved spaces. The results presented in this paper are a step towards bridging this gap, and clarifying their usefulness and power in analyzing observed data. Careful analysis of any observed data to identify physical origins of anisotropy and non-Gaussianity, if any, using TMFs requires that we acquire deeper understanding of the mathematical theory and implications of TMFs for random field. 
$W_2^{1,1}$ contains additional information of the statistical nature (Gaussian or not) of random fields in comparison to $W_1$. In two dimensions, the two first derivatives of the field, $u_{;1}, u_{;2}$, are independent of each other. Since the two eigenvalues of $W_2^{1,1}$ are integrals over the square of each $u_{;i}, i=1,2$, they provide independent probes of the Gaussian nature or deviation from it, of the field.
The tensor generalization of the genus,  $W_2^{0,2}$, must always have identical eigenvalues, as seen in section 2.1.  However, for random fields the eigenvalues are again integrals over the square of each independent field derivative. Therefore, they also provide independent probes of the Gaussian nature of the field. In summary, the translation invariant rank-2 TMFs are advantageous over the scalar MFs primarily due to two reasons. The first is that they provide new morphological information. Secondly, they expand the suit of real space based statistical quantities that can be used to constrain non-Gaussianity of cosmological fields - from the traditional scalar MFs. It is worth mentioning that it is possible to devise statistics that can capture some shape information using suitable combinations of the scalar MFs, such as the so-called `shape finders'~\cite{Varun:1998} that have been used to quantify filamentarity of the large scale structure.  

%What is done in the paper
We have generalized the definition of the translation invariant rank-2 TMFs to the sphere. We have obtained analytic formulae for the ensemble expectations of the TMFs for the special cases where the field is isotropic Gaussian and isotropic Rayleigh. The analytic formulae are valid on flat space as well as curved space. We clarify how the statistical isotropy and the nature of the fields are encoded in the TMFs and consequently in the alignment parameter $\alpha$.  This method can be applied to cosmological fields to search for alignments in the field which may be of physical origin or due to spurious effects arising from contaminations of the fields or due to instrumental effects. We then apply the method to synchrotron, thermal dust, anomalous microwave and CO  emissions from our Galaxy given by the PLANCK 2015 data release as examples of fields that demonstrate high level of alignment.

% What can be extended and ongoing
It is straightforward to extend our calculations to TMFs on 3-dimensional flat space for application to fields such as matter and galaxy distributions. We are addressing this in our forthcoming work.

%What more to do with foregrounds
The analysis of the PLANCK foreground emissions presented here is sketchy since our purpose was only to show examples of fields that exhibit alignment of structures. Our results do however reveal very interesting non-Gaussian and statistical anisotropy features of the different fields. It would be useful to follow up this line of investigation in order to gain insights into the distribution and properties of the material that give rise to the foreground emissions. We plan to carry this out in the near future.

%Ongoing applications
The application of TMFs to cosmological fields promises to be very fruitful. As mentioned in the introduction, the first application to the CMB data from PLANCK was carried out in~\cite{Vidhya:2016}. The authors have found that $E$ mode data shows significant departure from statistical isotropy.
A re-analysis of the cleaned PLANCK data to search for any residual/anomalous alignments using the method developed in this paper is ongoing. Applications of the method to the large scale structure to study the growth of structure and probing statistical isotropy are forthcoming. Applications to the fields of the epoch of reionization can also reveal interesting physical effects.

% What we don't address and can be extended
This paper does not address the analytic calculation of TMFs for individual curves and hence does not address the calculation of $\beta$.
For an isotropic random field, even though we obtain $\alpha=1$ implying that the average over all  curves is circular, individual curves have intrinsic anisotropy. For Gaussian isotropic CMB temperature field it was shown in \cite{Vidhya:2016} that the average anisotropy is quantified by $\beta \sim 0.62$. It may be possible to compute the probability distribution of $\beta$ for isotropic Gaussian fields by the use of conditional probability.
Further, this paper does not calculate TMFs separately for hotspots and coldspots since the method used here is not adequate for this computation. The results of  \cite{Vidhya:2016} suggest that $\alpha$ varies with the threshold differently for hotspots and coldspots. It would be useful to extend our calculation and obtain analytic forms for TMFs of hotspots and coldspots separately.

%%%%%%%%%%%%%%%%%%%%%%%%%%%%%%%%%%%%%%%%%%%%%%%%%%%%%%%%%%%%%%%%%%%%%%%%%%%%%%
\section*{Acknowledgment}{We acknowledge the use of the \texttt{Hydra} cluster at the Indian Institute of Astrophysics. We also acknowledge use of the \texttt{QUEST} cluster at the Center for Advanced Computation, Korea Institute for Advanced Study, for preliminary calculations which helped gain some insights. Some of the results in this paper have been obtained by using the \texttt{CAMB}~\cite{Lewis:2000ah,cambsite}  and \texttt{HEALPIX}~\cite{Gorski:2005,Healpix} packages.
We acknowledge use of observational data obtained with Planck, an ESA science mission with instruments and contributions directly funded by ESA Member States, NASA, and Canada. P.C would like to thank the Korea Institute of Advanced Studies for a visiting professorship during which some of this work was initiated. P.C would also like to thank Joseph Samuel for useful discussions on integral geometry. P.C and Joby P. K. would also like to thank Tuhin Ghosh for discussion about the CMB foreground component separation by the PLANCK team.  Joby P. K. would like to thank Ravikumar C. D. for his help at Calicut University.} 

%%%%%%%%%%%%%%%%%%%%%%%%%%%%%%%%%%%%%%%%%%%%%%%%%%%%%%%%%%%%%%%%%%%%%%%%%%%%%%
\appendix

\section{Definition of tensorial integration}
\label{sec:appen}

The definitions here follow~\cite{Fabian:1957}. Let $V^{m,n}$ be an $(m,n)$-rank tensor defined at all points of a curve $\mathcal C$ on a $d$-dimensional space $M$ with an affine connection. Let $V^{i_1,...,i_m}_{j_1,...,j_n}$ be the components of $V^{m,n}$ in the coordinate system $(x^1,...,x^d)$. Consider the set of $(m+n)$ differential equations
\begin{equation}
V^{i_1,...,i_m}_{j_1,...,j_n} = \frac{D}{D\lambda}X^{i_1,...,i_m}_{j_1,...,j_n} 
\end{equation}
where  $\frac{D}{D\lambda}$ is the covariant derivative along the curve parametrized by parameter $\lambda$. If these differential equations have a solution 
\begin{equation}
X^{i_1,...,i_m}_{j_1,...,j_n} = Y^{i_1,...,i_m}_{j_1,...,j_n}, 
\end{equation}
then $Y^{i_1,...,i_m}_{j_1,...,j_n}$ is called the tensor integral of $V^{i_1,...,i_m}_{j_1,...,j_n}$ along the curve with respect to $\lambda$, and is expressed as
\begin{equation}
Y^{i_1,...,i_m}_{j_1,...,j_n} = \int_{\mathcal C} \, V^{i_1,...,i_m}_{j_1,...,j_n}\,{\rm d} \lambda. 
\end{equation}
By definition the tensor $Y^{i_1,...,i_m}_{j_1,...,j_n}$ has the same transformation properties as $V^{i_1,...,i_m}_{j_1,...,j_n}$.   
%%%%%%%%%%%%%%%%%%%%%%%%%%%%%%%%%%%%%%%%%%%%%%%%%%%%%%%%%%%%%%%%%%%%%%%%%%%%%%

\section{Some basic formulae}
\label{sec:basicmath}

A smooth curve in 2-dimensions can be defined parametrically as a vector
\begin{equation}
\vec r(t)=x(t) \,\hat {\rm i} + y(t) \,\hat {\rm j},
\end{equation}
where $t$ is the parameter, and $x(t),\, y(t)$ are differentiable functions of $t$. The tangent vector to the curve at any point $t$ is given by
\begin{equation}
\vec T \equiv \frac{{\rm d}\vec r}{{\rm d}t} = \frac{\dx}{\dt} \,\hat {\rm i} + \frac{{\rm d} y}{\dt} \,\hat {\rm j},
\end{equation}
The length of the curve between two points indexed by $t_1$ and $t_2$ is given by
\begin{equation}
  s  = \int_{t_1}^{t_2} \,\ds \ =\ %\int_{t_1}^{t_2} \,\sqrt{ \left(\frac{\dx}{\dt}\right)^2 + \left(\frac{{\rm d} y}{\dt}\right)^2}\, \dt=
  \int_{t_1}^{t_2} \,|\vec T| \, \dt,
\end{equation}
where $\ds$ is the infinitesimal arc length. 
The curvature, $\kappa$, at any point of the curve is a measure of how quickly the tangent vector turns with respect to the arc length. It is given by 
\begin{equation}
\kappa  = \bigg|\frac{{\rm d}\hat T}{\ds}\bigg| = \frac{1}{|\vec T|}\,\bigg|\frac{{\rm d}\hat T}{\dt}\bigg|.
\end{equation}
%In terms of derivatives with respect to $t$,
%$\dot x,\ \ddot x,\ \dot y, \ \ddot y$,
$\kappa$ can be expressed as
\begin{equation}
\kappa  = \frac{|\dot x \ddot y -\dot y\ddot x|}{(\dot x^2+\dot y^2)^{3/2}},
\end{equation}
where the dot represents derivative with respect to $t$. 
The {\em signed curvature} is assigned positive sign for counterclockwise curves and negative for clockwise curves.
The unit normal vector, $\hat n$, which is the vector normal to the tangent vector at any point of the curve is given by 
\begin{equation}
\hat n  = \frac{1}{\big|\frac{{\rm d}\hat T}{\dt}\big|} \frac{{\rm d}\hat T}{\dt}.
\end{equation}
This formula fails for straight lines where the tangent vector is constant.

Using these expressions it is straightforward to show that for an ellipse with $p$ and $q$ as the semi-major and minor axes, respectively, and the $x$ and $y$-axes chosen along the semi-major and minor axes,
$W_2^{1,1}$ is given by the following analytic expression,
\begin{equation}
  W_2^{1,1}= \left( \begin{array}{cc} f_2^{1,1}(p,q) & 0 \\ 0 & f_2^{1,1}(q,p), \end{array} \right),
  \label{eqn:W211formula1}
\end{equation}
where
\begin{equation}
f_2^{1,1}(p,q)=\frac{1}{2} p^2 q^2 \int_0^{2\pi} {\rm d} \varphi \frac{cos^2 \varphi}{\left[ p^2-(p^2-q^2)cos^2 \varphi \right]^{3/2}}.
\label{eqn:W211formula2}
\end{equation}
The definition of $W_2^{1,1}$ can be re-expressed as
\begin{equation}
  \int \vec{r} \otimes \hat{n} \,\kappa \,\ds = %\int \vec{r} \otimes \frac{d\hat{T}}{dt} \,dt =
  \int \frac{d\vec r}{dt} \otimes \hat{T} \,\dt -  \int {\rm d}\, \big[ \vec r \otimes \hat{T}\big]  = \int \, \hat T \otimes \hat{T} \,\ds.
\end{equation}
In the last line we have used $\int d\, \big[ \vec r \otimes \hat{T}\big] =0$. This proves Eq. (11).  
Next,
\begin{equation}
  {\mathbf {Tr}} \int \hat{T} \otimes \hat{T}  \ds =  \int {\mathbf {Tr}} \, \bigg(\hat{T} \otimes \hat{T}\bigg) \ds =\int \bigg(\hat{T}_1^2 + \hat{T}_2^2 \bigg) \ds = \int \ds
  \label{eqn:traceproof1}
\end{equation}
\begin{equation}
  {\mathbf {Tr}} \int \hat{T} \otimes \hat{T}\, \kappa\, \ds =  \int {\mathbf {Tr}}\, \bigg(\hat{T} \otimes \hat{T} \bigg)\, \kappa\, \ds =\int \bigg(\hat{T}_1^2 + \hat{T}_2^2 \bigg) \,\kappa\, \ds = \int \kappa \,\ds,
\label{eqn:traceproof2}
\end{equation}
where we have used the fact that integration and taking the trace commute since they are both linear. This holds in flat as well as curved spaces. Eqs. (\ref{eqn:traceproof1}) and  (\ref{eqn:traceproof2}) are the proofs of Eqs. (\ref{eqn:tmf_trace1}) and  (\ref{eqn:tmf_trace2}).

Inserting $\vec n=\nabla u$ and Eq. (19) in Eq. B5 we obtain
\begin{equation}
  \kappa = \frac{ 2u_{;1}u_{;2}u_{;12} - u_{;1}^2 u_{;22} - u_{;2}^2 u_{;11}}{{|\nabla u|^3}},
\end{equation}
where ; represents covariant derivative for curved space. This proves Eq. (20).
%%%%%%%%%%%%%%%%%%%%%%%%%%%%%%%%%%%%%%%%%%%%%%%%%%%%%%%%%%%%%%%%%%%%%%%%%%%%%%
\section{Proof of positivity of eigenvalues of $\mathcal{W}_1$}

We can discretize the loop integral of $W_2 ^{1,1}$ defined in Eqn. (11) and write the entries of the matrix  in the form
\begin{equation}
  {\mathcal{W}}_1=\left(
  \begin{array}{cc}
  \sum_i (T_1(i) )^2 & \sum_i (T_1(i) T_2(i))\\
  \sum_i(T_1(i)T_2(i)) & \sum_i (T_2(i))^2\\
  \end{array}\right)
\end{equation}
Here, the labels $1,2$ refer to the components of the unit tangent vectors.

The eigenvalues are then easily determined to be positive using the
Cauchy-Schwarz inequality $\sum_i(T_1(i) )^2 \sum_j (T_2(j) )^2
>(\sum_i T_1(i)T_2(i) )^2 $

In the case of $\overline{\mathcal W}_1$, the label $i$ which runs over a
single discretized loop now simply needs to be augmented to a pair of
summation indices $i,n$ where the second label $n$ refers to the sum
over loops. The positivity again follows from the Cauchy-Schwarz 
inequality.

%%%%%%%%%%%%%%%%%%%%%%%%%%%%%%%%%%%%%%%%%%%%%%%%%%%%%%%%%%%%%%%%%%%%%%%%%%%%%%

 %%%%%%%%%%%%%%%%%%%%%%%%%%%%%%%%%%%%%%%%%%%%%%%%%%%%%%%%%%%%%%%%%%%%%%%%%%%%
\end{document}